\documentclass{elsart}
\usepackage{graphicx}
\usepackage{amssymb}

\begin{document}

\begin{frontmatter}



\title{Novel scaling behavior of the Ising model on curved surfaces}


\author{I. Hasegawa, Y. Sakaniwa and H. Shima}

\address{Department of Applied Physics, Graduate School of Engineering, Hokkaido University, Sapporo, 060-8628 Japan}

\begin{abstract}
We demonstrate the nontrivial scaling behavior of Ising models defined on (i) a donut-shaped surface and (ii) a curved surface with a constant negative curvature.
By performing Monte Carlo simulations, we find that the former model has two distinct critical temperatures at which both the specific heat $C(T)$ and magnetic susceptibility $\chi(T)$ show sharp peaks.
The critical exponents associated with the two critical temperatures are evaluated by the finite-size scaling analysis; the result reveals that the values of these exponents vary depending on the temperature range under consideration.
In the case of the latter model, it is found that static and dynamic critical exponents deviate from those of the Ising model on a flat plane; this is a direct consequence of the constant negative curvature of the underlying surface.
\end{abstract}

\end{frontmatter}

% main text

\section{Introduction}

The recent progress in fine processing technologies has enabled the fabrication of nanoscale materials with novel shapes \cite{spiral,polygons,bull,colloidal,mobius,biomedicine,hollow,hollow2,spiral2}.
A study of the basic properties of these materials is crucially important for many potential applications; it also helps to gain a new perspective on nanodevice modeling.
From a fundamental point of view, it is interesting to determine how the geometry of these nanostructures influences their physical properties, particularly their critical behavior close to the phase transition.
It is known that at the phase transition, certain physical quantities exhibit power-law behavior that is universal depending on certain essential symmetries of the system \cite{Cardy,Fisher,NakayamaYakubo}.
Hence, an alteration in the geometric symmetry of the system possibly results in an alteration in critical behavior. 
On this background, several studies have suggested that systems with novel geometry exhibit peculiar critical behavior \cite{Diego,Hoelbling,Gonzalez,Weigel,Deng,Costa-Santos,Pleimling,Freitas,Moura-Melo}.
Nevertheless, no systematic investigation has been conducted on the relationship between geometry and critical phenomena that occur in magnetic nanostructures.

In this context, one of the most interesting issue is to study the critical properties of low-dimensional magnetic systems with nonzero geometric curvature.
The two-dimensional Ising model with ferromagnetic interaction is a typical example of a system capable of a magnetic phase transition.
It is well known that this model undergoes the ferromagnetic transition in zero external field as the temperature decreases below a critical temperature $T_c$ \cite{Yang,McCoy}.
In the vicinity of $T_c$, thermodynamic quantities exhibit the power-law behavior that is characterized by critical exponents.
The values of these exponents are determined only by certain global symmetries (and dimensionality) of the system; this has been confirmed both theoretically and experimentally in the case of the Ising model on a flat plane \cite{Onsager,Kaufman,Yeomans,Ma}.
Currently, we are interested in the effect of geometry on the critical exponents, that is, whether a geometric alteration induces a quantitative alteration in the values of the critical exponents of the system.

In this study, we investigate the critical behavior of the Ising models defined on two types of curved surfaces: (i) a donut-shaped surface produced by gluing a flat plane and (ii) a surface with a constant negative curvature (i.e., a pseudosphere).
Monte Carlo simulations combined with finite-size scaling analysis reveal that in both Ising models, the value of the critical exponent $\gamma$ of the zero-field magnetic susceptibility $\chi(T)$ (as well as the dynamic critical exponent $z$) deviates from that of the conventional Ising model on a flat plane.
Furthermore, the donut-shaped Ising lattice model has two distinct critical temperatures at which the specific heat and magnetic susceptibility show sharp peaks.
These surprising results indicate that the geometry of the underlying surface is significantly related to the phase transition of low-dimensional magnetic materials.

%%%%%%%%%%%%%%%%%%%%%%%%%%%%%%%%%%%%%%%%%%%%%%%%%%%%
\begin{figure}[ttt]
\begin{center}
\includegraphics[width=0.4\textwidth]{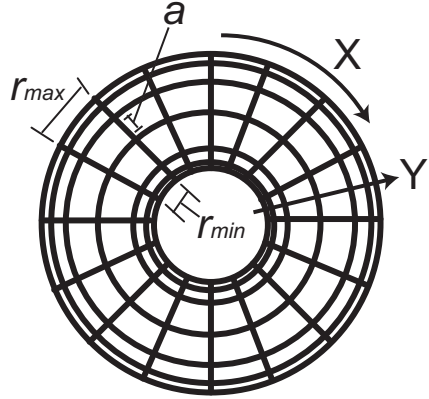}
\end{center}
\caption{Schematic illustration of the square lattice defined on a donut-shaped surface obtained by gluing the edges of the flat plane.
The spacial modulation of the separation occurs only along the X direction.}
\label{fig:donut}
\end{figure}
%%%%%%%%%%%%%%%%%%%%%%%%%%%%%%%%%%%%%%%%%%%%%%%%%%%%

\section{Donut-shaped Ising models}

First, we consider a square Ising lattice model defined on a donut-shaped surface (see Fig.~\ref{fig:donut}) \cite{Hasegawa}.
The donut-shaped lattice is constructed by gluing the edges of a square lattice on a flat plane; hence, this lattice is {\it topologically} equivalent to a square flat lattice with the periodic boundary condition.
However, this equivalence breaks down when the difference in the {\it intrinsic geometry} between the flat plane and the donut-shaped surface is considered. 
The gluing procedure does not preserve the separation $r_{ij}$ of the neighboring sites $i$ and $j$, thereby resulting in the spatial modulation of the coupling strength $J_{ij}$ of the nearest-neighbor spins $s_i$ and $s_j$.
To our knowledge, it has not yet been clarified how the gradual modulation of $J_{ij}$ affects the thermodynamic properties of the embedded Ising model. 
(In contrast, the Ising models consisting of random or alternative bonding have been extensively studied thus far \cite{Honecker,Carter,Amoruso}.)

The donut-shaped Ising model is described by the Hamiltonian 
\begin{equation}
\mathcal{H} = -\sum_{i,j}J_{ij}{s_i}{s_j} \quad \mbox{with} \quad J_{ij} \equiv \frac{J_0a}{r_{ij}},
\end{equation}
where the constants $a$ and $J_0$ represent the separation of neighboring spins and the coupling constant of the square lattice before the gluing procedure, respectively (in other words, these constants belong to the square lattice on the flat plane).
In the definition of $J_{ij}$, the effect of the spatial modulation of $r_{ij}$ is taken into account so that $J_{ij}$ becomes spatially dependent.
In the actual calculation, we have constructed donut-shaped Ising lattices with different sizes by gluing planar square lattices with linear dimensions of $L$ = 200, 250, and 300 in units of $a$.
For all donut-shaped lattices, the minimum and maximum separations are fixed as $r_{\rm min} = 0.88$ and $r_{\rm max} = 2.88$, respectively (see Fig.~\ref{fig:donut}).

%%%%%%%%%%%%%%%%%%%%%%%%%%%%%%%%%%%%%%%%%%%%%%%%%%%%
\begin{figure}[ttt]
\begin{center}
\includegraphics[width=0.55\textwidth]{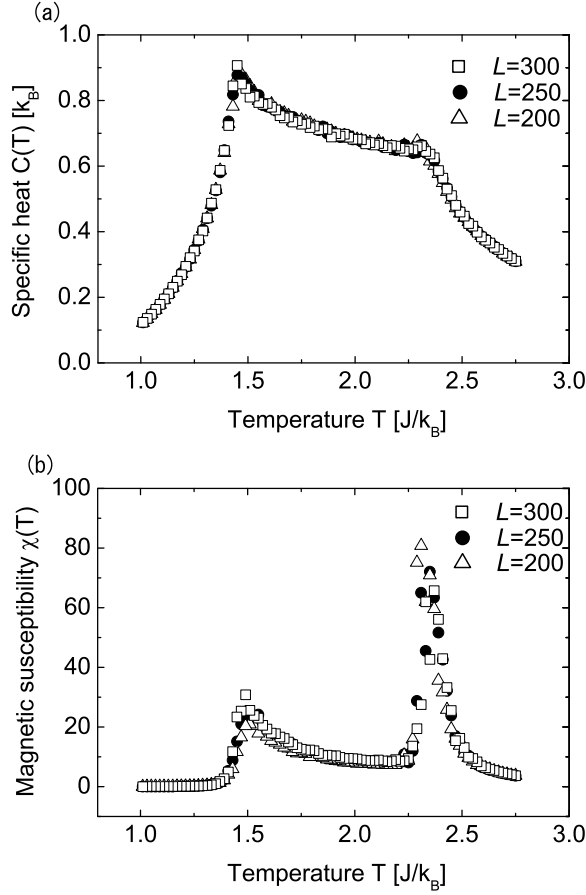}
\end{center}
\caption{(a) Specific heats and (b) magnetic susceptibilities of the donut Ising model.
There appear two sharp peaks at $T_{c1} = 1.47$ and $T_{c2} = 2.40$.}
\label{fig:spechi}
\end{figure}
%%%%%%%%%%%%%%%%%%%%%%%%%%%%%%%%%%%%%%%%%%%%%%%%%%%%

\section{Occurrence of two critical temperatures}

We have performed Monte Carlo simulations \cite{BinderHeermann} to evaluate the specific heats $C(T)$ and  magnetic susceptibilities $\chi(T)$ defined by
\begin{equation}
C(T) = \frac{\langle E^2 \rangle - \langle E \rangle^2}{k_BT^2},
\end{equation}
and
\begin{equation}
\chi(T) = \frac{\langle M^2 \rangle - \langle M \rangle^2}{k_BT}.
\end{equation}
Here $E$ and $M$ are the internal energy and the spontaneous magnetization of the system, respectively; the angular brackets indicate to take thermal average.
Figure~\ref{fig:spechi} shows the numerical results for various system size; $L$=200 (open triangles), 250 (solid circles), and 300 (open squares).
Most striking is the fact that both data exhibit two sharp peaks at $T_{c1} = 1.47$ and $T_{c2} = 2.40$ in unit of $J_0/k_B$.
The values of these critical temperatures are well apart from that of the square Ising model on a flat plane \cite{BinderHeermann}:
\begin{equation}
T_c = \frac{2}{\ln(1+\sqrt{2})} \sim 2.27. 
\end{equation}
We have also attempted to evaluate the critical exponents of $\chi(T)$ by means of the finite-size scaling analysis. The scaling hypothesis states that in the vicinity of $T_c$, $\chi$ for a finite system size $L$ obeys the scaling form \cite{Cardy}:
\begin{equation}
\chi(T,L) \propto L^{{\gamma}/{\nu}}f(|T-T_c|L^{{1}/{\nu}}),
\end{equation}
where $\gamma$ and $\nu$ are referred to as the critical exponent of $\chi(T)$ and that of the correlation length $\xi(T)$  of the order parameter, respectively.
Intriguingly, in our donut Ising model, the value of the exponent $\gamma$ depends strongly on the temperature range. For instance, at immediately lower and higher than $T_{c1}$, the exponent yields $\gamma \sim 3.0$ and $\gamma \sim 0.6$, respectively.
On the other hand, it yields $\gamma \sim 0.9$ at just higher than $T_{c2}$. 
These values of $\gamma$ obviously deviate from that for the planar Ising model: $\gamma_{2d} = 7/4$.

%%%%%%%%%%%%%%%%%%%%%%%%%%%%%%%%%%%%%%%%%%%%%%%%%%%%
\begin{figure}[ttt]
\begin{center}
\includegraphics[width=0.7\textwidth]{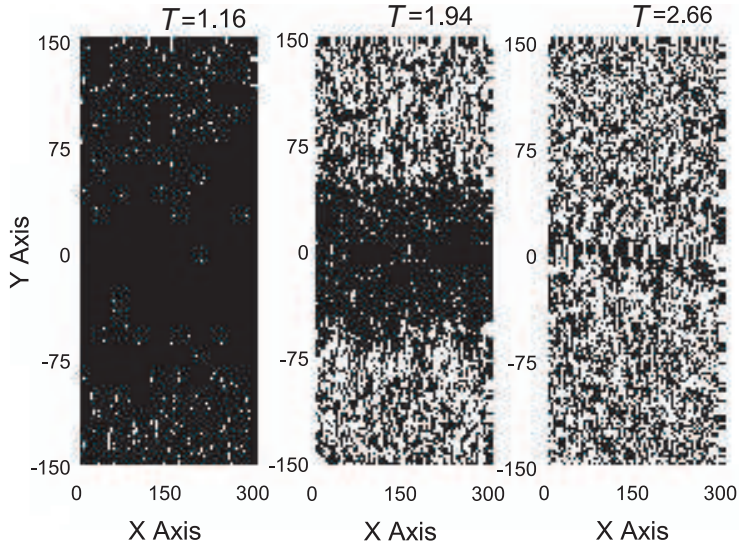}
\end{center}
\caption{Spatial configurations of Ising spins on the donut surface.
Black and white dots represent the positive and negative sign of the Ising spin, respectively.
Each axis corresponds to the direction of $X$ and $Y$ on the donut surface depicted in Fig.~\ref{fig:donut}.
The horizontal lines $Y=0$ and $Y= \pm 150$ represent the internal and external circumference of the donut surface.}
\label{fig:config}
\end{figure}
%%%%%%%%%%%%%%%%%%%%%%%%%%%%%%%%%%%%%%%%%%%%%%%%%%%%

The occurrence of the two distinct critical temperatures in the donut-shaped Ising model is qualitatively explained by observing the spatial configurations of the Ising spins $s_i$ at different temperatures.
Figure~\ref{fig:config} shows the spatial configurations of the Ising spins on the donut surface; black and white dots represent the positive and negative signs of the Ising spin, respectively.
At $T > T_{c2}$ (in the right panel), the spin directions are completely random and therefore the system has no spontaneous magnetization.
On the contrary, at $T < T_{c1}$ (in the left panel), most spins are oriented in the same direction; hence, the system has long-range order.
The configuration in the intermediate temperature range of $T_{c1} < T < T_{c2}$ (in the middle panel) is important.
In the last case, a large ferromagnetic domain exists along the internal circumference of the donut-shaped surface ($Y \sim 0$), while the region close to the external circumference ($Y \sim 150$ or $-150$) remains in a paramagnetic state.

The partial growth of the ferromagnetic domain results from the gradual modulation of $J_{ij}$ over the donut-shaped surface.
Close to the internal circumference, the magnitude of $J_{ij}$ is sufficiently large to overcome the thermal fluctuation; hence, the ferromagnetic domain at the inner part persists until the temperature reaches $T_{c2}$.
On the other hand, the ferromagnetic order along the external circumference disappears at $T_{c1}$, since $J_{ij}$ in the latter region is very weak.
Consequently, the two critical temperatures are produced by local ferromagnetic transitions near the internal and external circumferences of the donut-shaped surface.

It is emphasized that the peculiar behaviors of the donut-shaped Ising model, such as the two peaks of $\chi(T)$ and the quantitative shift of $\gamma$, are direct consequences of the gluing procedure; in other words, they are caused by the mechanical deformation of a flat magnetic plane.
Hence, these findings imply that it is possible to control the physical properties of magnetic nanostructures by inducing a local (or global) mechanical deformation.
The relevance of our results to actual magnetic systems with novel geometry is also under consideration.

\section{Ising models on negatively curved surfaces}

In the previous sections, we have considered the donut-shaped Ising lattice model in which the separation of neighboring spins is spatially dependent due to the external gluing procedure.
Alternatively, however, it is possible to construct the {\it regular} Ising lattice model on certain curved surfaces, wherein the Ising spins are assigned with an equiseparation over the entire surface.
A typical example is an Ising model defined on negatively curved surfaces, i.e., on a pseudosphere \cite{Wu,Doyon1}.
The pseudosphere is a simply connected infinite surface whose Gaussian curvature at any point has a constant negative value \cite{Coxeter,FirbyGardiner}.
More importantly, in the latter model, the surface curvature effects manifest themselves in scaling behavior in a different way from that in the donut-shaped Ising model.
In what follows, we will give a brief review of our previous study on the Ising model on the pseudosphere \cite{Shima,Shima2}. 
By comparing the results obtained for the two distinct curved surfaces, we gain a deeper understanding of the effect of geometry on the Ising model.

%%%%%%%%%%%%%%%%%%%%%%%%%%%%%%%%%%%%%%%%%%%%%%%%%%%%
\begin{figure}[ttt]
\begin{center}
\includegraphics[width=0.7\textwidth]{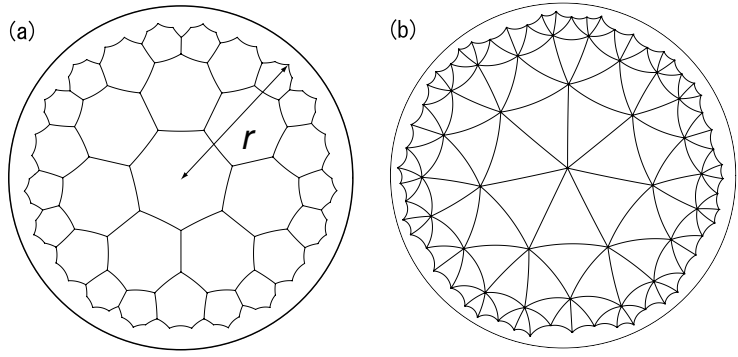}
\end{center}
\caption{Local bonding structures of regular (a) heptagonal and (b) triangular lattices in terms of the Poincar\'e disk representation.
The size of our lattice is determined by the number of concentric layers of polygons, denoted by $r$.}
\label{fig:poincare}
\end{figure}
%%%%%%%%%%%%%%%%%%%%%%%%%%%%%%%%%%%%%%%%%%%%%%%%%%%%

Figure \ref{fig:poincare} illustrates  (a) regular heptagonal and (b) regular triangular Ising lattices embedded on the pseudosphere in terms of the Poincar\'e disk representation \cite{Balazs}.
Although polygons depicted in the figures appear to be distorted, they all are surely congruent in the sense of the intrinsic geometry on the pseudosphere.
Onto these lattices, we assign the ferromagnetic Ising model with nearest neighbor interaction, described by the Hamiltonian
\begin{equation}
\mathcal{H} = -J \sum_{i,j} s_i s_j ,
\end{equation}
and perform Monte Carlo simulations \cite{Binder} to evaluate the critical exponent $\gamma$ of the magnetic susceptibility $\chi$.
It is noted that in the present case, the scaling form of $\chi$ should be expressed as
\begin{equation}
\chi(T,N) \propto N^{\gamma/\mu} \cdot \chi_0 \left( |T-T_{\rm c}|N^{1/\mu} \right),
\end{equation}
where $\mu$ describe the divergence of the correlation volume as $\xi_{\rm V}(T)\propto |T-T_{\rm c}|^{-\mu}$.
Similar to the case of the donut Ising model, quantitative evaluation of these exponents can be achieved by using the finite-size scaling technique \cite{Shima,Shima2,Binder}.

Before addressing numerical results, it deserves comment on the boundary contribution in the Ising model on the pseudosphere.
It follows from Fig.~\ref{fig:poincare} that the size of our lattice is determined by the number of concentric layers of polygons, denoted by $r$.
When $r\gg 1$, the number of total sites is approximated as $N(r) \propto e^r$;  this exponential increase in $N$ means that the contribution from the boundaries survives even in the thermodynamic limit $r = \infty$.
In order to extract the bulk critical properties, therefore, we have taken into account only the Ising spins involved in the interior $r_{\rm in}$ layers ($r_{\rm in} \le r_{\rm out}=r_{\rm in}+\delta r$) when performing the scaling analysis.
Asymptotic behaviors of $\gamma$ and $\mu$ for large $\delta r$ (as well as large $r_{\rm in}$) provide estimations of the bulk critical exponents.

%%%%%%%%%%%%%%%%%%%%%%%%%%%%%%%%%%%%%%%%%%%%%%%%%%%%
\begin{figure}[ttt]
\begin{center}
\includegraphics[width=0.95\textwidth]{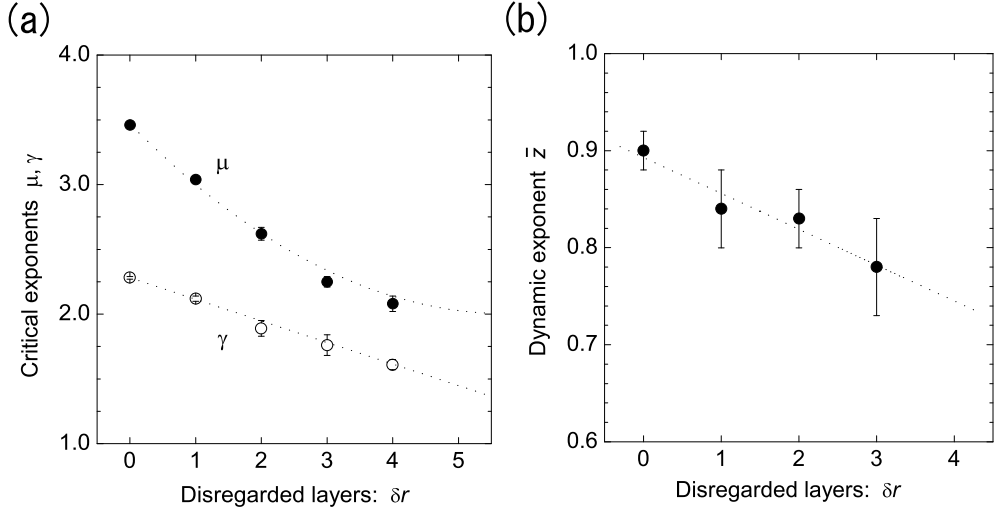}
\end{center}
\caption{$\delta r$-dependences of (a) $\gamma$ and $\mu$ and (b) $\bar{z}$.
Each data point was extracted by means of the finite size scaling analysis for the system sizes $4\le r_{\rm in} \le 8$.}
\label{fig:results}
\end{figure}
%%%%%%%%%%%%%%%%%%%%%%%%%%%%%%%%%%%%%%%%%%%%%%%%%%%%

Figure~\ref{fig:results} (a) shows the values of critical exponents $\mu$ and $\gamma$ as a function of the number of disregarded layers $\delta r$.
We see that, while the curve of $\mu$ converge to a particular value of $\mu \sim 2$ or less, that of $\gamma$ has no tendency to converge to $\gamma_{\rm 2d}=7/4$ for large $\delta r$.
Notably, these results are consistent in quality with the conjecture deduced by the quantum-field theory \cite{Doyon2}.
It states that the Ising model embedded on the pseudosphere should yield the mean-field critical exponents (i.e., $\gamma_{\rm MF}=1$ and $\mu_{\rm MF}=2$) when the boundary contribution may be omitted.
These mean-field behaviors can be simply attributed to the relation $N \propto N_s$ between $N$ and the number of spins along the boundary $N_s$ \footnote{Since the relation $N_s \propto N^{1-(1/d)}$ holds for an Ising lattice in $d$ dimension, $N_s \propto N$ effectively consequences $d= \infty$.}, which implies that the pseudosphere act as an infinite-dimensional surface.

It is also interesting to evaluate the dynamic critical exponent $\bar{z}$ of the Ising model on the pseudosphere, which can be done by using the short-time relaxation method \cite{Janssen,Soares}.
The $\delta r$-dependence of $\bar{z}$ are summarized in Fig.~\ref{fig:results} (b).
We observe that $\bar{z}$ monotonically decreases with $\delta r$; this indicates that for sufficiently large $\delta r$, $\bar{z}$ takes a value considerably smaller than that of the planar Ising models: $\bar{z} = z/d \simeq 1.1$.
That is, $\bar{z}$ on the pseudosphere is also expected to exhibit the mean field exponent $\bar{z}_{\rm MF}=z_{\rm MF}/d_c=1/2$ at $\delta r \gg 1$, since $z_{\rm MF}=2$ for the Ising model and $d_c=4$ the upper critical dimension.
Quantitative determination of $\bar{z}$ for $\delta r \gg 1$ as well as those of other static exponents (such as $\beta$ and $\eta$) are under consideration; preliminary results are given in Ref. \cite{Sakaniwa}.

\section{Conclusion}

To summarize, we have investigated the critical behaviors of the two-dimensional Ising models embedded on a donut-shaped surface and a pseudosphere.
In the case of the donut-shaped Ising model, the specific heat $C(T)$ and magnetic susceptibility $\chi(T)$ exhibit two sharp peaks at $T_{c1} = 1.47$ and $T_{c2} = 2.40$.
The two critical temperatures are produced by the spatial modulation of the coupling strength $J_{ij}$ that leads to the formation of a partial ferromagnetic domain along the internal circumference of the donut-shaped surface.
Furthermore, we have attempted to evaluate the critical exponent $\gamma$ of $\chi(T)$.
When the temperature is less than $T_{c1}$, the exponent yields $\gamma \sim 3.0$; however, when the temperature is greater than $T_{c1}$, it yields $\gamma \sim 0.6$; further, at a temperature just above $T_{c2}$, it yields $\gamma \sim 0.9$.
Evidently, these values of $\gamma$ deviate from those of the planar Ising model.
We have also demonstrated that in the case of the Ising model on the pseudosphere, the static exponent $\gamma$ and the dynamic exponent $\bar{z}$ deviate from those of the planar Ising model.
We believe that the results of our study provide  new information for the theoretical development of critical phenomena that occur on general curved surfaces.

\ack 
We thank T. Nakayama and K. Yakubo for stimulating discussions.
This work was supported by a Grant-in-Aid for ScientificResearch from the Japan Ministry of Education, Science, Sports and Culture.
H.S thanks the financial support from the Mazda Foundation and the Murata Science Foundation.

% \appendix

% \section{}
% \label{}

\end{document}